\documentclass{article}
\usepackage{times}
\usepackage{amsfonts}
\usepackage{graphicx}
\begin{document}
\hyphenation{Min-kows-kian}
\noindent
{\Large SUMMING OVER SPACETIME DIMENSIONS\\ IN QUANTUM GRAVITY}\\

\vskip1cm
\noindent
{\bf Erik Curiel}$^{1,a}$, {\bf Felix Finster}$^{2,b}$ and {\bf J.M. Isidro}$^{3,c}$\\
$^{1}$Munich Center for Mathematical Philosophy, Ludwig-Maximilians-Universit\"at,\\ Geschwister-Scholl-Platz 1, D-80539 M\"unchen, Germany\\
$^{2}$Fakult\"at f\"ur Mathematik, Universit\"at Regensburg, D-93040 Regensburg, Germany\\
$^{3}$Instituto Universitario de Matem\'atica Pura y Aplicada,\\ Universidad Polit\'ecnica de Valencia, Valencia 46022, Spain\\
$^{a}${\tt Erik.Curiel@lrz.uni-muenchen.de}, $^{b}${\tt finster@ur.de},\\ $^{c}${\tt joissan@mat.upv.es}
\vskip.5cm
\noindent
\today
\vskip.5cm

\noindent
{\bf Abstract}  Quantum--gravity corrections (in the form of a minimal length) to the Feynman propagator for a free scalar particle in $\mathbb{R}^D$ are shown to be the result of summing over all  dimensions $D'\geq D$ of $\mathbb{R}^{D'}$, each summand taken in the absence of quantum gravity.

\section{Introduction}\label{suno}

Feynman propagators encode more information than meets the eye. Let us for simplicity consider a massive scalar particle on a $D$--dimensional manifold ${\cal M}$. When the latter is Minkowskian flat space $\mathbb{R}^D$, it has been found \cite{PADDY2, PADDY4, PADDY5}  that inertial scalar propagators  suffice in order to derive the thermal properties of the Rindler horizon. A slight modification of Schwinger's proper--time representation of the scalar propagator turns out to provide an ultraviolet completion of the scalar theory, both in flat spaces \cite{PADDY0, PADDY1} and in spaces of constant nonzero curvature \cite{PADDY3}; this ultraviolet completion amounts to the existence of a quantum of length $L$. These are just a few examples of (in principle unexpected) quantum--gravity properties of Feynman scalar propagators, {\it the latter considered in the absence of gravity}\/.

In the naive path--integral approach to quantum gravity one has to sum over all metrics on the given manifold ${\cal M}$. In the presence of several inequivalent differentiable structures and/or topologies one has to sum over them: one integrates over more than one manifold structure. This raises the question of summing over all possible dimensions as well. 

In this letter we consider a free scalar particle in the spacetime ${\cal M}=\mathbb{R}^D$, the latter endowed with its standard topology, differentiable structure, and Euclidean or Minkowskian metric, and perform a sum of scalar Feynman propagators over all dimensions $D'\geq D$. Each one of these summands is free of quantum--gravity effects; {\it but the sum of all summands will turn out to produce quantum--gravitational effects on the scalar particle}\/. This will provide yet another unexpected example of quantum--gravity effects that are encoded within flat--spacetime, inertial Feynman propagators. As will be explained in more detail below, the quantum--gravity effects under consideration are those arising from the existence of a quantum of length $L$.

We present our analysis first in Euclidean $\mathbb{R}^D$ in section \ref{sdos}. In section \ref{stres} we Wick--rotate back to Minkowskian $\mathbb{R}^D$, the latter with signature $(+,-,\ldots, -)$. In section \ref{dappy} we report an alternative derivation of the same results due to Padmanabhan (private communication). In our use of Bessel functions we follow the conventions and notations of ref. \cite{GRADSHTEYN}. We use natural units throughout.

\section{Quantum--gravitational properties of inertial propagators: Euclidean space}\label{sdos}

\subsection{An ultraviolet completion}

A massive scalar particle in Euclidean momentum space $\mathbb{R}^D$ has the Feynman propagator 
$(p^2+m^2)^{-1}$. Fourier transforming and inserting Schwinger's proper--time integral
\begin{equation}
\frac{1}{p^2+m^2}=\int_0^{\infty}{\rm d}s\,\exp\left[-s(p^2+m^2)\right],
\label{trece}
\end{equation}
one arrives at
\begin{equation}
G_D(r)=\frac{m^{D-2}}{(2\pi)^{D/2}}\frac{K_{D/2-1}\left(mr\right)}{\left(mr\right)^{D/2-1}}, \qquad r^2:=\sum_{j=1}^Dx_j^2.
\label{catorce}
\end{equation}
Above, $K_\nu(z)$ is a modified Bessel function, a solution to the modified Bessel equation $z^2f''(z)+zf'(z)-(z^2+\nu^2)f(z)=0$. The latter possesses two linearly independent solutions, conventionally denoted $I_{\nu}(z)$ and $K_{\nu}(z)$. For later use we recall two well--known properties \cite{WATSON}. First, the $K_{\nu}(z)$ are singular at $z=0$ while the $I_{\nu}(z)$ are everywhere regular. Second, the $I_n(z)$ have the following generating function:
\begin{equation}
\sum_{n=-\infty}^{\infty}I_n(z)\,t^n=\exp\left[\frac{z}{2}\left(t+\frac{1}{t}\right)\right].
\label{once}
\end{equation}

Now the right--hand side of (\ref{once}) can be regarded as a certain ultraviolet completion of the standard integral kernel for the Feynman propagator, the latter expressed as a path integral. Indeed, in refs. \cite{PADDY0, PADDY1} one modifies the standard path integral
\begin{equation}
G_D(x)=\sum_{\rm paths}\exp\left[-mS(x)\right],
\label{treintayuno}
\end{equation}
where the action integral $S(x)=\int^x{\rm d}s$ equals the proper length, to become 
\begin{equation}
G_D^{{\rm (QG)}}(x)=\sum_{\rm paths}\exp\left\{-m\left[S(x)+\frac{L^2}{S(x)}\right]\right\}.
\label{treintaydos}
\end{equation}
The superindex QG in the corrected Feynman propagator (\ref{treintaydos}) stands for {\it quantum--gravity}\/. It has been argued in refs. \cite{PADDY0, PADDY1} that the propagator (\ref{treintaydos}) includes the effects of the presence of a quantum of length as given by the minimal  length $L$. In this sense, the modified path integral (\ref{treintaydos}) can be regarded as an ultraviolet completion of the standard scalar theory on Euclidean space.

The right--hand side of the modified path integral (\ref{treintaydos}) closely resembles the generating function (\ref{once}), while reducing to the usual case (\ref{treintayuno}) in the limit $L\to 0$. Indeed the result of computing the path integral (\ref{treintaydos}),
\begin{equation}
G_D^{\rm (QG)}(r)=\frac{m^{D-2}}{(2\pi)^{D/2}}\frac{K_{D/2-1}\left(m\sqrt{r^2+4L^2}\right)}{\left(m\sqrt{r^2+4L^2}\right)^{D/2-1}}
\label{siete}
\end{equation}
correctly reduces to the propagator (\ref{catorce}) when $L=0$. Moreover, when $L\neq 0$, this ultraviolet completion of the scalar Euclidean propagator has the property that {\it it no longer diverges as $r\to 0$}\/, because the presence of the quantum of length $L$ prevents it. So quantum--gravity effects mollify the Feynman propagator at short distances, while at the same time ensuring invariance under the transformation
\cite{PADDY0, PADDY1}
\begin{equation}
S\longrightarrow\frac{L^2}{S}.
\label{sesenta}
\end{equation}

We close this section with some observations. The ultraviolet completion of the Euclidean scalar propagator summarised above (as originally presented in refs. \cite{PADDY0, PADDY1}) was {\it not}\/ based on the generating--function--for--Bessel--functions approach followed here. Rather, it was based on the requirement of invariance under the duality (\ref{sesenta}). However, we have found it useful to reword it in the language of generating functions, with an eye on what comes next.

\subsection{An identity satisfied by the $K_\nu(z)$}

Do the Bessel functions $K_n(z)$ also satisfy an identity of the type (\ref{once})? To the best of our knowledge, no such expression has been published in the standard literature \cite{GRADSHTEYN, WATSON}. In what follows we derive a new identity satisfied by the $K_n(z)$. It reads
\begin{equation}
\sum_{n=0}^{\infty}\frac{t^{n}}{n!}K_{n+\nu}(2z)=\left(\frac{z}{\sqrt{z^2-zt}}\right)^{\nu}K_\nu\left(2\sqrt{z^2-zt}\right), \quad \vert\arg(z)\vert<\frac{\pi}{4}.
\label{cuarenta}
\end{equation}
In order to prove Eq. (\ref{cuarenta}) we start from the integral representation \cite{GRADSHTEYN}
\begin{equation}
K_\mu(z)=\frac{1}{2}\left(\frac{z}{2}\right)^\mu\int_0^{\infty}{\rm d}s\,\exp\left(-s-\frac{z^2}{4s}\right)s^{-\mu-1}, \quad \vert\arg(z)\vert<\frac{\pi}{4}
\label{cincuenta}
\end{equation}
and consider the series
\begin{equation}
f_{\nu}(z,t):=\sum_{n=0}^{\infty}\frac{1}{n!}K_{n+\nu}(2z)\,t^{n},\qquad t\in\mathbb{C}.
\label{cincuentayuno}
\end{equation}
Substituting (\ref{cincuenta}) into the above and rearranging terms we find
$$
f_{\nu}(z,t)=\frac{1}{2}\int_0^{\infty}\frac{{\rm d}s}{s}\exp\left(-s-\frac{z^2}{s}\right)\sum_{n=0}^{\infty}\frac{1}{n!}\left(\frac{zt}{s}\right)^{n+\nu}t^{-\nu}
$$
$$
=\frac{1}{2}\int_0^{\infty}\frac{{\rm d}s}{s}\exp\left(-s-\frac{z^2-zt}{s}\right)\left(\frac{z}{s}\right)^{\nu}=\left(\frac{z}{\sqrt{z^2-zt}}\right)^{\nu}K_\nu\left(2\sqrt{z^2-zt}\right),
$$
which establishes (\ref{cuarenta}).

\subsection{Summing over dimensions}

As was already the case with the known identity (\ref{once}), our new identity (\ref{cuarenta}) will turn out to have an interesting application to quantum gravity. We first evaluate (\ref{cuarenta}) at $z=y/2$ and $t=-1/(2y)$ to obtain
\begin{equation}
\sum_{n=0}^{\infty}\frac{(-1/2)^n}{n!}\frac{K_{n+\nu}(y)}{y^{n}}=\left(\frac{y}{\sqrt{y^2+1}}\right)^{\nu}K_\nu\left(\sqrt{y^2+1}\right), 
\label{cuarentayuno}
\end{equation}
valid whenever $y\neq 0$ and  $\vert\arg(y)\vert<\pi/4$.  Without loss of generality it will be convenient to set 
\begin{equation}
m=1,\qquad L=1/2
\label{noventa}
\end{equation}
when solving Eq.~(\ref{catorce}) for the Bessel functions $K_{\mu}$ in terms of Euclidean Feynman propagators $G_{D}$:
\begin{equation}
K_{\mu}\left(r\right)=(2\pi)^{\mu+1}r^{\mu}G_{2\mu+2}(r), \qquad \mu=\frac{D}{2}-1,
\qquad r^2=\sum_{j=1}^{2\mu+2}x_j^2.
\label{dieciseis}
\end{equation}
Next we set $y$ on the right--hand side of (\ref{cuarentayuno}) equal to $r=\left(\sum_{j=1}^{2\nu+2}x_j^2\right)^{1/2}$, bearing in mind that the same $y$ will appear on the left--hand side. We remark that the natural radial variable that the summand $K_{n+\nu}(y)$ on the left--hand side of  (\ref{cuarentayuno}) should depend on is not $\left(\sum_{j=1}^{2\nu+2}x_j^2\right)^{1/2}$ but $\left(\sum_{j=1}^{2n+2\nu+2}x_j^2\right)^{1/2}$; we will return to this point presently. Substitution of Eq. (\ref{dieciseis}) into (\ref{cuarentayuno}) finally gives
\begin{equation}
\sum_{n=0}^{\infty}\frac{(-\pi)^n}{n!}G_{2n+2\nu+2}(r)=G_{2\nu+2}\left(\sqrt{r^2+1}\right)=G^{{\rm (QG)}}_{2\nu+2}(r)
,\quad r^2=\sum_{j=1}^{2\nu+2}x_j^2.
\label{cincuentaydos}
\end{equation}

We know that quantum--gravity effects on the Euclidean scalar propagator in $D=2\nu+2$ dimensions cause the appearance of a quantum of length $L$. These effects have been taken into account in Eq. (\ref{cincuentaydos}), as shown by the right--hand side. The left--hand side expresses this quantum--gravitationally corrected propagator as an {\it infinite sum of gravity--free propagators}\/. Each summand corresponds to one higher value of the dimension $2n+2\nu+2$, one for each $n\in\mathbb{N}$. Within each $(2n+2\nu+2)$--dimensional space that contributes to the above sum, however, only a $(2\nu+1)$--dimensional sphere $S^{2\nu+1}\subset\mathbb{R}^{2n+2\nu+2}$ is swept out by the equation $r={\rm const}$. With increasing values of the dimension $2n+2\nu+2$ it is always this same sphere $S^{2\nu+1}$ that is swept out, {\it i.e.}\/, the sphere does not lie along the additional dimensions. Hence the higher dimensions being summed over in Eq. (\ref{cincuentaydos}) play the role of a {\it virtual}\/ spacetime for propagation in the {\it actual}\/ dimension $2\nu+2$. Being virtual, however, does not imply that they are unphysical, as they add up to a nonvanishing quantum--gravitational correction to the Feynman propagator in $2\nu+2$ dimensions.

\section{Quantum--gravitational properties of inertial propagators: Minkowski space}\label{stres}

$D$--dimensional Minkowskian and Euclidean propagators are related as per
\begin{equation}
G^{(M)}_D(t, {\bf x})=-{\rm i}G^{(E)}_D({\rm i}\tau, {\bf x}), \qquad {\bf x}\in\mathbb{R}^{D-1},
\label{cien}
\end{equation}
the superindices $M,E$ referring to Minkowski and Euclidean space, respectively. By Eq. (\ref{catorce}) one thus finds
\begin{equation}
G^{(M)}_D(t, {\bf x})=\frac{m^{D/2}}{2^{D+1}\pi^{D/2-1}{\rm i}^D}\frac{H^{(2)}_{D/2-1}\left(m\sqrt{t^2-{\bf x}^2}\right)}{\left(m\sqrt{t^2-{\bf x}^2}\right)^{D/2-1}},
\label{cientouno}
\end{equation}
where $H_\mu^{(2)}=J_\mu-{\rm i}Y_\mu$ is a Hankel function of the second kind \cite{GRADSHTEYN}. With the understanding that we will henceforth work in Minkowski space, we will drop the superindex $M$ from our notations. Also, for simplicity we will restrict our attention to timelike vectors and denote
\begin{equation}
r^2:=t^2-{\bf x}^2=t^2-\sum_{j=1}^{D-1}x_j^2,
\label{cientrocuatro}
\end{equation}
so $r>0$. In the units of Eq. (\ref{noventa}), the gravity--free propagator (\ref{cientouno}) simplifies to
\begin{equation}
G_D(r)=\frac{1}{2^{D+1}\pi^{D/2-1}{\rm i}^D}\frac{H^{(2)}_{D/2-1}\left(r\right)}{r^{D/2-1}}.
\label{cientoveinte}
\end{equation}

The quantum--gravity corrected Feynman propagator in Minkowski space $\mathbb{R}^D$ is readily obtained from the above: following refs. \cite{PADDY0, PADDY1}, it suffices to make the replacement $r=\sqrt{r^2}\rightarrow \sqrt{r^2+4L^2}$. This yields in (\ref{cientouno})
\begin{equation}
G^{\rm (QG)}_D(r)=\frac{m^{D/2}}{2^{D+1}\pi^{D/2-1}{\rm i}^D}\frac{H^{(2)}_{D/2-1}\left(m\sqrt{r^2+4L^2}\right)}{\left(m\sqrt{r^2+4L^2}\right)^{D/2-1}},
\label{cientoseis}
\end{equation}
which, in the units of Eq. (\ref{noventa}), becomes
\begin{equation}
G^{\rm (QG)}_D(r)=\frac{1}{2^{D+1}\pi^{D/2-1}{\rm i}^D}\frac{H^{(2)}_{D/2-1}\left(\sqrt{r^2+1}\right)}{\left(\sqrt{r^2+1}\right)^{D/2-1}}.
\label{cientosiete}
\end{equation}
Again, the role of the quantum of length is to mollify the singularity of the propagator at the origin.

We now proceed to show that the effects of including a quantum of length $L$ can be equivalently obtained as a sum over all dimensions $D'$ higher than the given dimension $D$, all summands in the absence of gravity. Following the same reasoning as in the Euclidean case, first we need an identity similar to (\ref{cuarenta}) for the Hankel functions $H^{(2)}_\mu$. Happily, Eqs. (5) and (13) on p. 141 of the standard reference \cite{WATSON} provide us with the sought--for identity:
\begin{equation}
\sum_{n=0}^{\infty}\frac{\left(-\frac{1}{2}tz\right)^n}{n!}H^{(2)}_{\nu+n}(z)=\left(1+t\right)^{-\nu/2}H^{(2)}_{\nu}\left(z\sqrt{1+t}\right),\quad \vert t\vert<1.
\label{cientodos}
\end{equation}
Next we solve Eq. (\ref{cientoveinte}) for the Hankel functions $H^{(2)}$ in terms of the gravity--free Feynman propagators $G$, and substitute the result into Eq. (\ref{cientodos}). With $z=r$ in the latter, this produces
\begin{equation}
\sum_{n=0}^{\infty}\frac{(2\pi tr^2)^n}{n!}G_{2\nu+2n+2}(r)=\left(1+t\right)^{-\nu/2}G_{2\nu+2}\left(\sqrt{r^2+tr^2}\right),\quad \vert t\vert<1.
\label{cientocinco}
\end{equation}
Let us now set $tr^2=1$. Then $\vert t\vert<1$ will hold provided that $r>1$, and (\ref{cientocinco}) reads
\begin{equation}
\sum_{n=0}^{\infty}\frac{(2\pi)^n}{n!}G_{2\nu+2n+2}(r)=\left(1+\frac{1}{r^2}\right)^{-\nu/2}G_{2\nu+2}\left(\sqrt{r^2+1}\right),\quad r>1
\label{cientodiez}
\end{equation}
which for very large $r$ becomes
\begin{equation}
\sum_{n=0}^{\infty}\frac{(2\pi)^n}{n!}G_{2\nu+2n+2}(r)=G_{2\nu+2}\left(\sqrt{r^2+1}\right)=G^{{\rm (QG)}}_{2\nu+2}\left(r\right),\quad r>>1.
\label{cientoonce}
\end{equation}
We recall that the variable $r$ on both sides of the above equation is defined through
\begin{equation} 
r^2=t^2-\sum_{j=1}^{2\nu+1}x_j^2.
\label{cientotreinta}
\end{equation}

It should be observed that one {\it cannot}\/ derive Eq. (\ref{cientoonce}) from a Wick rotation of Eq. (\ref{cincuentaydos}); the reason is twofold. Not only does a factor of $2\pi$ in (\ref{cientoonce}) replace a factor of $-\pi$ in (\ref{cincuentaydos}). More importantly, the sum over Euclidean dimensions (\ref{cincuentaydos}) is based on the identity (\ref{cuarenta}). The latter requires $\arg(z)<\pi/4$, a condition which is violated under multiplication by ${\rm e}^{\pm{\rm i}\pi/2}$. Although the Feynman propagators themselves can be Wick--rotated, the identity satisfied by the corresponding Bessel functions cannot. In other words: the operations of Wick rotation and ultraviolet completion do not commute.

\section{An alternative derivation due to Padmanabhan}\label{dappy}

After the first version of this paper appeared in the arxiv, we learnt from Padmanabhan (private communication) that he has obtained this result \cite{one} around the time he did the work in path integral duality by an alternative approach. His derivation is as follows: The Schwinger propertime representation of the QG-corrected propagator, in D-dimensional Euclidean sector, is given by the integral:
$$
G^{\rm (QG)}_D(r)=\int_0^\infty\frac{{\rm d}s}{(4\pi s)^{D/2}}\exp\left(-m^2s-\frac{r^2}{4s}\right)\exp\left(-\frac{L^2}{4s}\right)
$$
\begin{equation}
\equiv \int_0^\infty\frac{{\rm d}s}{(4\pi s)^{D/2}}F(s,r^2)\exp\left(-\frac{L^2}{4s}\right),
\end{equation} 
with $r^2$ as in Eq. (\ref{catorce}). Note that the  dependence on the dimension arises only from the factor $s^{-D/2}$. We now expand the factor $e^{-L^2/4s}$ in the integrand in a Taylor series. The $n$-th term in the series will introduce  the factor $s^{-n}$, which will change the original $s^{-D/2}$ factor to $s^{-(D+2n)/2}$, which occurs in the standard (uncorrected) propagator $G_{D+2n}(r)$ for the dimension $D+2n$. So we immediately get:
$$
G^{\rm (QG)}_D(r)=\sum_{n=0}^\infty\frac{\left(-\pi L^2\right)^n}{n!}\int_0^\infty\frac{{\rm d}s}{(4\pi s)^{(D+2n)/2}}F(s,r^2)
$$
 \begin{equation}
=\sum_{n=0}^\infty\frac{\left(-\pi L^2\right)^n}{n!} G_{D+2n}(r).
 \end{equation} 
The  QG corrected propagator can therefore be expressed as the sum of standard propagators for the dimensions $(D+2n)$. The simplicity of this derivation is noteworthy. It is obvious that the approach also works in the Lorentzian sector.

\section{Conclusions}\label{scuatro}

Our conclusions are summarised by Eqs.  (\ref{cincuentaydos}) and (\ref{cientoonce}), again collected below for convenience: in Euclidean 
$\mathbb{R}^D$ we have
\begin{equation}
G^{{\rm (QG)}}_{2\nu+2}(r)=\sum_{n=0}^{\infty}\frac{(-\pi)^n}{n!}G_{2n+2\nu+2}(r), \qquad D=2\nu+2
\label{doscientos}
\end{equation}
whereas in Minkowskian $\mathbb{R}^D$ we have
\begin{equation}
G^{{\rm (QG)}}_{2\nu+2}\left(r\right)=\sum_{n=0}^{\infty}\frac{(2\pi)^n}{n!}G_{2\nu+2n+2}(r),\qquad r>>1.
\label{trescientos}
\end{equation}
In both cases we find that a quantum--gravity corrected propagator in $D$ dimensions can be expressed as an infinite sum over all gravity--free propagators in dimension $D'\geq D$. Under {\it quantum--gravity corrections}\/ we understand, as already explained, the ultraviolet completion obtained by the inclusion of a minimal length \cite{PADDY0, PADDY1}. In the latter papers,  the existence of a quantum of length $L$ has been shown to be  equivalent to the requirement of invariance under the duality (\ref{sesenta}), the dimensionality of spacetime being kept fixed. The equivalent viewpoint that emerges from our analysis is that a smallest distance $L$ results from summing over an infinite number of higher dimensions.

Ultraviolet completions of standard theories have been the subject of many analyses, too numerous to quote here; see however ref. \cite{CROWTHER} for a sample of different standpoints.  As it turns out, UV completions are related to quantum gravity. We do not claim equivalence between these two issues. However, the duality symmetry (\ref{sesenta}) that implements our particular UV completion turns out to be equivalent to the existence of a quantum of length; {\it this is a crucial link between quantum gravity and UV completions}\/.

In this letter we have concentrated on the case of a massive scalar in flat $\mathbb{R}^D$, either Euclidean or Minkowski. The particular ultraviolet completion studied here, Eq. (\ref{treintaydos}), is usually regarded as enforcing the duality (\ref{sesenta}). Our approach in this letter interprets this same completion on the basis of the generating function (\ref{once}) for the modified Bessel functions $I_n(z)$ (and similar identities satisfied by their close cousins the Macdonald functions $K_n(z)$, the Hankel functions $H^{(2)}(z)$, etc.)
Alternatively but equivalently, these identities (Eqs. (\ref{once}), (\ref{cuarenta}) and (\ref{cientodos})) amount to a sum over an infinite number of virtual dimensions. On the basis of the aforementioned identities we have derived the expansions (\ref{doscientos}) and (\ref{trescientos}) for Feynman propagators. 

Even in flat space there is a remnant of quantum gravity, through the presence of a quantum of length $L$. The very existence of $L$ is a zeroth--order, quantum gravity  effect. Of course there are higher--order corrections due to curvature; but even flat space can feel the existence of a quantum of length. This quantum is commonly expected to equal the Planck length $L_P$. Now $L_P$ can be expressed in terms of $G$, $c$ and $\hbar$, all of which are perfectly well defined in flat space. This additional fact supports the statement that flat space still bears some imprint of quantum gravity.

Altogether, our analysis somehow places the dimensionality of spacetime on an equal footing with other variables that are integrated over in quantum gravity, such as the metric and the topology of the spacetime manifold ${\cal M}$. That the notion of dimensionality might not be as fundamental as believed has also been hinted at in various settings such as string theory \cite{BLAU} and others. It is an intriguing question to ask, if our expansions (\ref{doscientos}) and (\ref{trescientos}) could possibly find a thermal analogue in the fact that thermal scalar Green functions can be written as an infinite, imaginary--time sum of the corresponding zero--temperature Green functions \cite{PADDY4}. 

There is an interesting physical interpretation to our sum over dimensions. Namely, the sharply--defined concept of dimension in classical spacetime might be subject to some sort of quantum uncertainty once one enters the realm of quantum gravity. We can foresee an analogy with Heisenberg's principle of uncertainty: classical trajectories become diffuse in quantum theory, all possible classical trajectories are summed over in the Feynman path integral. It also points toward {\it a possible interpretation of the notion of dimension in thermodynamical terms}\/: some values of the dimension might be more densely populated than others, and the sharply--defined classical dimension might emerge as some sort of thermal average. We hope to report in the future.

\vskip1cm
\noindent
{\bf Acknowledgments} The research of E.C. was funded by grant CU 338/1-1 from Deutsche Forschungsgemeinschaft (Germany). The research of J.M.I. was supported by grant no. RTI2018-102256-B-I00 (Spain). The authors acknowledge  support from Vielberth Stiftung, Regensburg (Germany).

\end{document}